\documentclass[conference]{IEEEtran}
\IEEEoverridecommandlockouts
\usepackage{cite}
\usepackage{jabbrv}
\usepackage{url}
\usepackage{amsmath,amssymb,amsfonts}
\usepackage{algorithmic}
\usepackage{graphicx, epstopdf}
\usepackage{float}
\usepackage{caption}
\usepackage{subcaption}
\usepackage{textcomp}
\usepackage{xcolor}
\usepackage{svg}
\usepackage{longtable}
\usepackage{comment}
\usepackage[numbers,sort&compress]{natbib}
\def\BibTeX{{\rm B\kern-.05em{\sc i\kern-.025em b}\kern-.08em
    T\kern-.1667em\lower.7ex\hbox{E}\kern-.125emX}}
\usepackage{multirow}
\usepackage[normalem]{ulem}
\useunder{\uline}{\ul}{}
\usepackage[linesnumbered,commentsnumbered,ruled,vlined,norelsize, linesnumbered, ruled, lined, boxed, commentsnumbered]{algorithm2e}

\newcommand{\proj}
{\textit{Hermes}\xspace}

\begin{document}

\title{When Less is More: Achieving Faster Convergence in Distributed Edge Machine Learning}

\author{
    {\centering
        Advik Raj Basani, Siddharth Chaitra Vivek, Advaith Krishna, Arnab K. Paul
    }\\
    {\it DaSHLab - Department of Computer Science and Information Systems}\\
    {\it BITS Pilani, K. K. Birla Goa Campus, India}\\
    {\small {\it \{f20221155, f20220569, f20212831, arnabp\}@goa.bits-pilani.ac.in}}\\
}

\maketitle

\begin{abstract}
        Distributed Machine Learning (DML) on resource-constrained edge devices holds immense potential for real-world applications. However, achieving fast convergence in DML in these heterogeneous environments remains a significant challenge. Traditional frameworks like Bulk Synchronous Parallel (BSP) and Asynchronous Stochastic Parallel (ASP) rely on frequent, small updates that incur substantial communication overhead and hinder convergence speed. Furthermore, these frameworks often employ static dataset sizes, neglecting the heterogeneity of edge devices and potentially leading to straggler nodes that slow down the entire training process. The straggler nodes, i.e., edge devices that take significantly longer to process their assigned data chunk – hinder the overall training speed. To address these limitations, this paper proposes \proj, a novel probabilistic framework for efficient DML on edge devices. This framework leverages a dynamic threshold based on recent test loss behavior to identify statistically significant improvements in the model's generalization capability, hence transmitting updates only when major improvements are detected, thereby significantly reducing communication overhead. Additionally, \proj employs dynamic dataset allocation to optimize resource utilization and prevents performance degradation caused by straggler nodes. Our evaluations on a real-world heterogeneous resource-constrained environment demonstrate that \proj achieves faster convergence compared to state-of-the-art methods, resulting in a remarkable $13.22$x reduction in training time and a $62.1\%$ decrease in communication overhead.
\end{abstract}

\section{Introduction}
Edge computing brings data storage and processing closer to the devices that generate it, unlike traditional models where all sensor and internet-of-things data is sent to a central server for processing. This shift towards edge computing is driven by the significant increase in the number and processing power of edge devices. Distributing compute improves application performance, reduces bandwidth requirements, and enables faster, near real-time insights.

Distributed Machine Learning (DML) is an architecture that addresses the ever-growing computational complexity of ML algorithms and the immense data required for model convergence. DML is broadly classified into two types: data parallel and model parallel~\cite{kim2016strads}. In data parallel training, the most commonly used approach, each node receives the entire model but only a portion of the data during each training round unlike model splitting in model parallel approach.

Traditionally, ML models were trained on powerful, resource-uniform servers in data centers. This uniformity ensured consistent training times per round across all nodes. However, the explosion of edge-based ML models, particularly for real-time inference, presents a challenge. Network overhead becomes significant when large data transfers occur between edge devices and distant data center servers.

To address these challenges, frameworks have been developed to shift computation from the cloud to the edge, leveraging the growing computational power and data generated at the edge~\cite{zhao2021federated}. However, running DML algorithms at the edge presents unique challenges. Limited memory and minimal resources on edge devices necessitate splitting datasets to respect these constraints. Additionally, edge clusters are inherently heterogeneous, i.e., devices have varying resource availability. Training time depends on the slowest device (straggler), and these stragglers can significantly impact convergence time. Unreliable networks with high latency and arbitrary node failures further complicate DML at the edge. Existing DML frameworks attempt to address the slowdown caused by communication overhead and heterogeneous devices, either by limiting the extent of communication~\cite{ho2013more} or the frequency of communication~\cite{loulergue2005bulk}, but at the cost of accuracy.

To mitigate the drawbacks of the state-of-the-art, we propose \proj\footnotemark \footnotetext{\proj refers to the Greek God of Speed, symbolic of faster convergence.}, a data-parallel, DML framework based on TensorFlow designed for heterogeneous edge environments. We make the following contributions in this work:
\begin{itemize}
    \item We introduce \proj\footnotemark, \footnotetext{Codebase for \proj: \color{blue}{https://github.com/DaSH-Lab-CSIS/Hermes}}a novel asynchronous DML framework over the Parameter Server (PS) architecture~\cite{li2014communication}, specifically designed for heterogeneous clusters.
    \item \proj tackles the challenge of heterogeneous devices by dynamically allocating batch sizes based on each device’s capabilities, improving training efficiency.
    \item Implementation of a communication-efficient strategy that identifies major changes in the generalization capability of a worker’s model and transmits only significant gradient updates, reducing network bandwidth usage and allowing workers to train independently.
    \item Introduction of a loss-based stochastic gradient descent (SGD) on the PS that effectively weights gradients from a worker for aggregation to the global model, based on how effectively the gradients can contribute to the global model’s convergence.
    \item Implementation of pre-fetching to minimize communication time between the PS and the workers, thereby maximizing the time available for training. Furthermore, to also address the memory limitations of edge devices, \proj incorporates model compression techniques to ensure broader applicability.
    \item Evaluation of \proj on a heterogeneous testbed and comparing with state-of-the-art algorithms shows upto $13.22$x reduction in training time and $62.1\%$ reduction in communication activity.
\end{itemize}

\section{Background and Related Work}

\begin{figure}[h]
    \centering
    \includegraphics[width=\columnwidth]{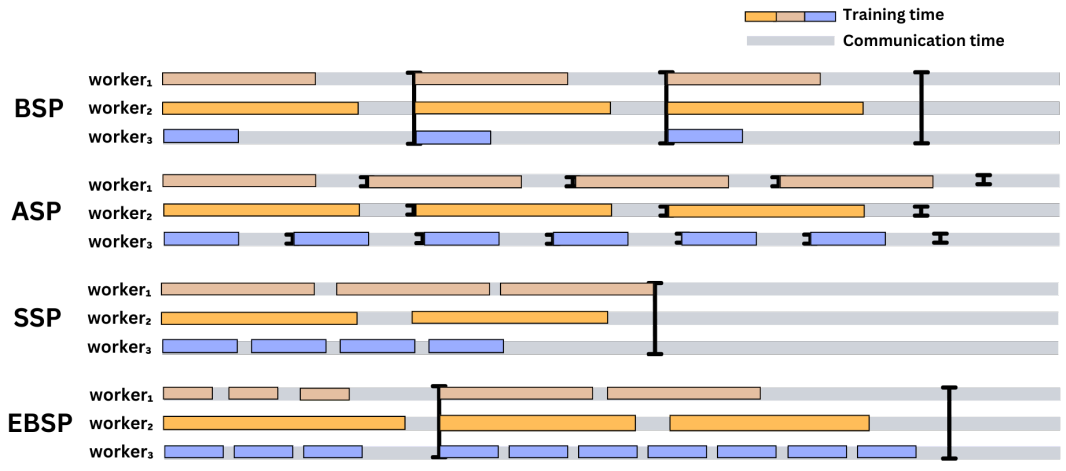}
    \hfill
    \caption{Training and communication time being displayed for BSP, SSP (s = 2, refer Sec.~\ref{sec:ssp}), ASP and EBSP; each line partition highlights a barrier where gradients of that node is pushed to the PS, with worker$_{2}$ and worker$_{3}$ as the slowest and fastest workers respectively.}
    \label{fig:all_sota}
\end{figure}

DML has become a cornerstone for training complex models by harnessing the combined processing power of multiple devices~\cite{dean2012large}. This parallel processing capability is crucial for handling the ever-growing volume of data used in modern DML tasks. Hence, efficient synchronization techniques are essential to ensure effective collaboration between these distributed computing units~\cite{li2014communication}. DML is often achieved by using a PS -- which is used for storing gradients, models, and datasets, and acts as a central server for nodes. In this system, one local epoch is defined as a node training on the entire dataset provided by the parameter server (PS) once, whereas a mini-batch refers to dividing the dataset into smaller sections, which are then used for training on a node. This section delves into various state-of-the-art (SOTA) synchronization paradigms used in DML, whose simplistic workflow can be seen in Fig. \ref{fig:all_sota}, including Bulk Synchronous Parallel (BSP)~\cite{loulergue2005bulk}, Asynchronous Parallel (ASP)~\cite{recht2011hogwild}, Stale Synchronous Parallel (SSP)~\cite{ho2013more}, Elastic BSP (EBSP)~\cite{zhao2019elastic} and Selective Synchronization (SelSync)~\cite{tyagi2023accelerating}. We will explore the specific advantages and limitations of each approach in terms of hardware efficiency and statistical convergence.

\subsection{Bulk Synchronous Parallel Method (BSP)}
BSP is a framework designed to enable practical DML, where each node operates on its own local memory, performing computations independently. Communication between processors occurs through a network, and barrier synchronization ensures that all processors align at a specific point called a superstep. During the local computation phase, workers train the model on their respective partitions of the dataset. Communication is done through a PS which partitions and sends the data to the workers as well as receives updates from them post-local training. Finally, the synchronization phase ensures that all updates from the workers are received before proceeding to the next superstep. Synchronous SGD (SyncSGD), in a model with parameter $w$, at the PS can be represented using Equation~\ref{eq1}, where $\eta$ represents the learning rate, $N$ denotes the number of workers, $g^{i}$ represents the gradients transmitted by worker $i$ at the superstep $j$ and with $w_{j}$ referring to values of $w$ at the j$^{th}$ iteration during aggregation:

\begin{equation}
\scriptsize
w_{j+1}^{i} = w_{j}^{i} + \eta\frac{1}{N} \sum_{i = 1}^{N} g^{i}
\label{eq1}
\end{equation}

This approach theoretically leads to faster convergence compared to mini-batch SGD~\cite{dean2012large}, but it comes at the cost of waiting for stragglers and increased communication overhead. 

\begin{figure}[t]
    \centering
    \includegraphics[width=5cm]{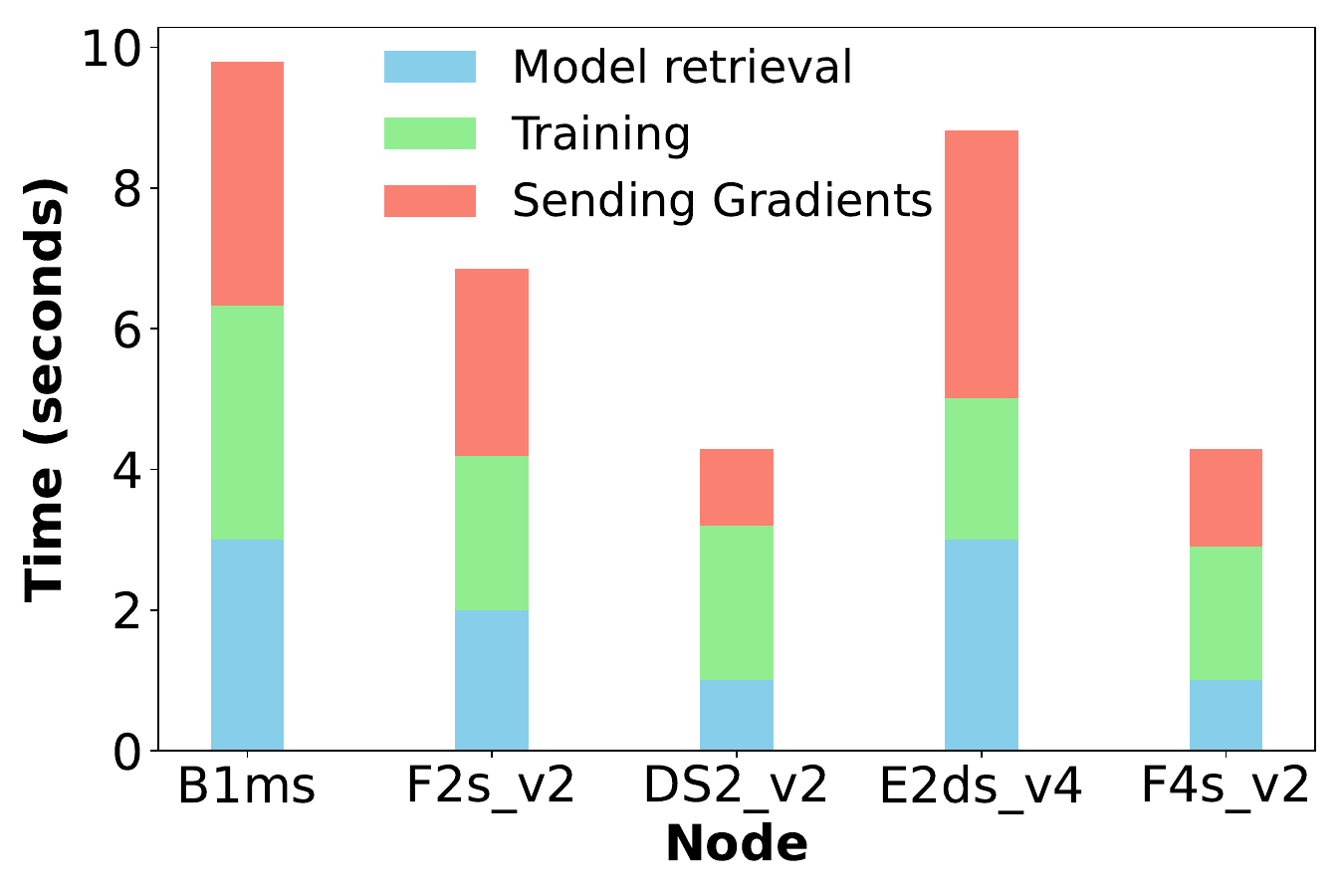}
    \caption{Comprehensive breakup of the time taken by workers during one local training cycle on MNIST dataset using CNN. The x-axis represents various node families to which our workers belong, refer to Table~\ref{table:4} for node configuration.}
    \label{fig:BSP_breakup}
\end{figure}

As a result, in context of edge computing, three components are essential to be optimized, which can viewed in Fig. \ref{fig:BSP_breakup}: the time taken for a worker to complete training, the time taken for receiving a partitioned dataset from the PS as well as receiving the global model after training, and the wait time required for all the workers to finish local training at the end of each superstep.

\subsection{Asynchronous Parallel Method (ASP)} \label{sec:asp}
In the context of ASP, workers are able to advance to subsequent iterations independently, without the need of waiting for all workers to complete their local training. This design choice ensures that no computational cycles are wasted, thereby achieving high levels of hardware efficiency. 

However, this approach comes with a significant trade-off in terms of statistical efficiency. Asynchronous SGD (AsyncSGD) can be mathematically represented as Equation~\ref{eq2}, using the same notation as SyncSGD:
\begin{equation}
\scriptsize
    w_{j+1}^{i} = w_{j}^{i} + \eta g^{i}
    \label{eq2}
\end{equation}

\begin{figure}[t]
    \centering
    \includegraphics[width=4.7cm]{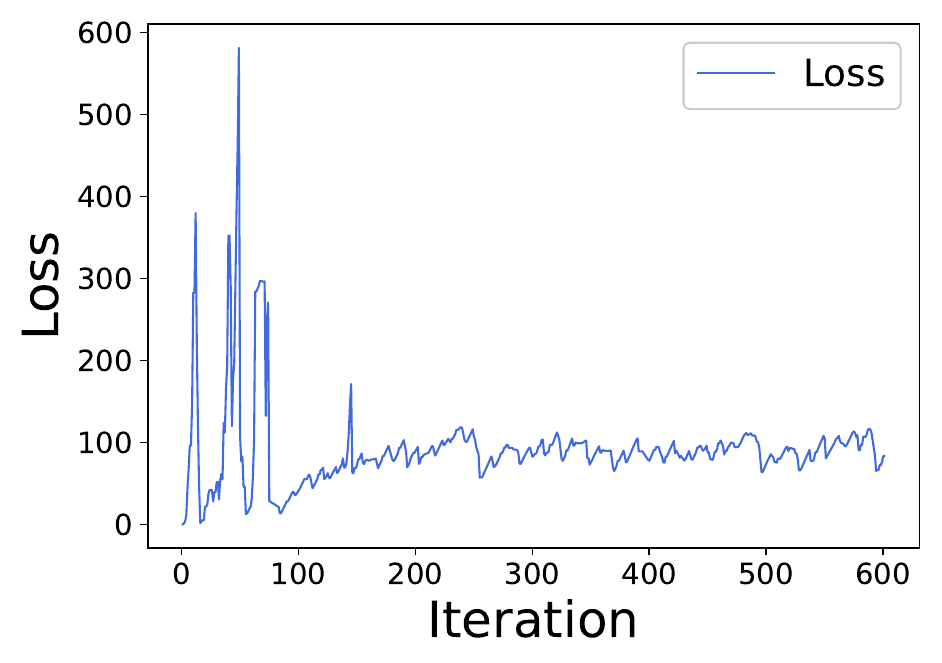}
    \caption{Loss fluctuations and failure to converge due to oscillation of global gradients in a cluster of nodes; cluster is trained on MNIST dataset using a CNN.}
    \label{fig:asp-sucks}
\end{figure}

Due to the asynchronous nature of updates, the direction of the gradient updates from different workers can vary significantly. When these updates are added subsequently during aggregation, they can cause the model parameters to oscillate back and forth instead of steadily converging toward the optimal solution. This is demonstrated in Fig. \ref{fig:asp-sucks}. The plot is indicative that the model is attempting to lower the loss and learn, but due to conflicting gradient directions, it fails to converge. Furthermore, due to the absence of global synchronization, the gradient computations performed by the workers are often based on outdated or stale parameter values. This leads to a decrease in the quality of the updates, as the parameters used in the gradient calculations don’t always reflect the most recent state of the model. As a result, the updates made to the model parameters are less effective, and hence a greater number of iterations are required to reach convergence compared to synchronous methods.

\subsection{Stale Synchronous Parallel Method (SSP)} \label{sec:ssp}
Stale Synchronous Parallel (SSP)~\cite{ho2013more} was introduced to reconcile the trade-offs between BSP and ASP. SSP allows worker nodes to proceed asynchronously with a controlled level of staleness, defined by a staleness threshold $s$. The staleness threshold indicates the maximum number of iterations a straggler can lag behind the fastest worker before forcing synchronization. This allows nodes to continue computing with slightly outdated parameters, thus avoiding unnecessary synchronization delays and maintaining a higher accuracy.

\subsection{Elastic Bulk Synchronous Parallel Method (EBSP)}
Motivated by the limitations of previous methods, EBSP aims to mitigate the strict synchronization requirement of BSP while maintaining its benefits through dynamic synchronization. The PS, based on recent training intervals and benchmarks, forecasts future training times to determine the optimal synchronization time. This prediction is constrained by a lookahead limit $R$, which in turn controls the algorithm's runtime~\cite{zhao2019elastic}.
EBSP’s elastic synchronization optimizes to find synchronization barriers where waiting times are minimized even if it means faster workers would finish training multiple times. By dynamically adjusting synchronization times and bounding them within the lookahead limit \(R\), EBSP achieves a balance between minimizing waiting time and ensuring convergence, without emulating the ASP model's risks. 

The main drawback of EBSP arises from the extra time needed to benchmark the nodes and the inherent design of algorithms like $Zipline$~\cite{zhao2019elastic}, which are more suitable for situations with stragglers. These algorithms do not perform as well when the hardware efficiency of all nodes is similar. 

\subsection{Selective Synchronization Parallel Method (SelSync)} \label{sec:selsync}
SelSync~\cite{tyagi2023accelerating} aims to balance parallel and statistical efficiency by alternating between synchronous and local-SGD (SGD on local worker models) updates based on the relative gradient change. Local-SGD updates enhance parallel efficiency by reducing iteration time, while synchronous updates are triggered when the relative gradient change exceeds a hyper-parameter $\delta$, ensuring consistent model states across workers despite higher communication overhead. Although relative gradient change looks like a promising approach, relative gradients are inherently noisy due to factors like stochastic gradients and mini-batching. This noise can lead to misleading changes in the metric, deeming it unreliable for making critical decisions.

SelSync also proposes a one-time shuffling operation before training, ensuring each worker has access to the entire dataset. This custom partitioning (SelDP) prevents skewed models and redundant processing of the same samples. However, this scheme is impractical for edge devices with limited memory as it requires storing the entire dataset in a shuffled order on each device.

\section{Motivation} \label{sec:moti}
\subsection{Need for major updates} \label{sec:major}
As models scale in complexity and datasets grow exponentially, the traditional approach of aggregating all updates from every node efficiently becomes increasingly difficult. This inefficiency is due to the substantial communication overhead associated with frequent updates to the PS. Consequently, there is a need to refine the update selection process to minimize this communication overhead.

Examining the limitations of BSP, ASP, SSP and EBSP makes it evident why selecting major updates is important.

\begin{figure}[t]
    \centering
    \begin{subfigure}{0.24\textwidth}
        \includegraphics[width=\textwidth]{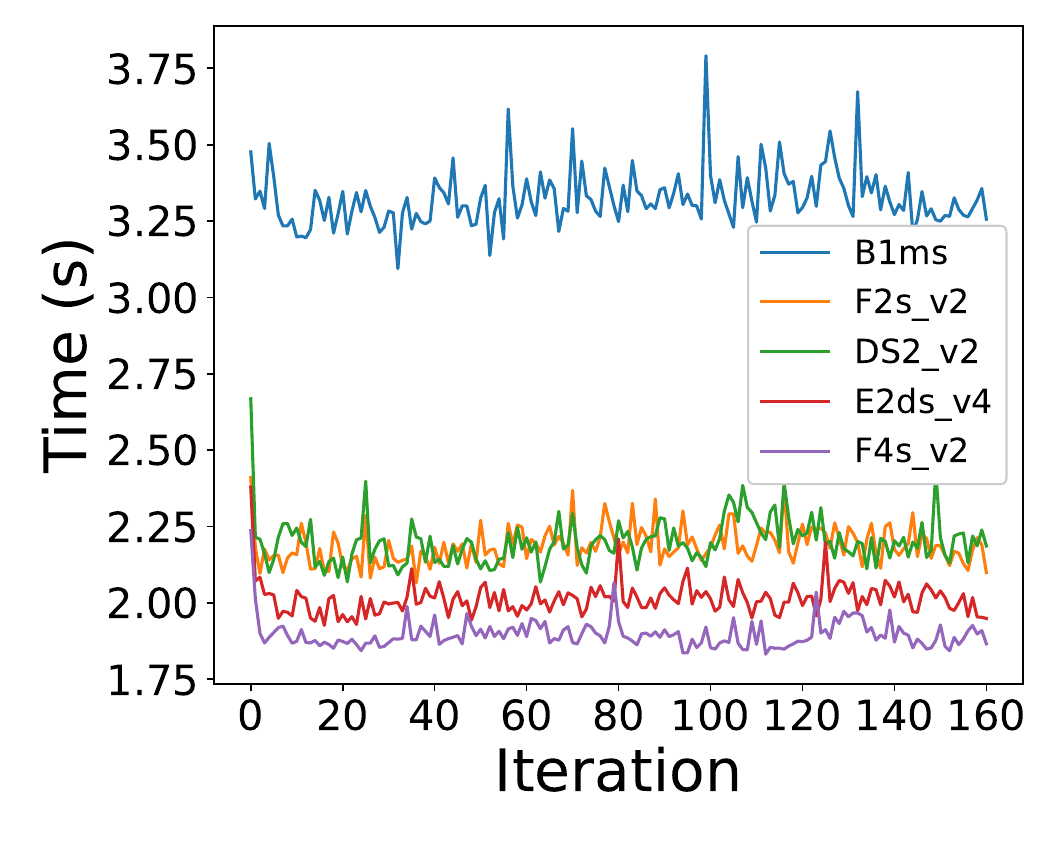}
        \caption{Training times per node.}
        \label{fig:node_training_times}
    \end{subfigure}
    \hfill
    \begin{subfigure}{0.24\textwidth}
        \includegraphics[width=\textwidth]{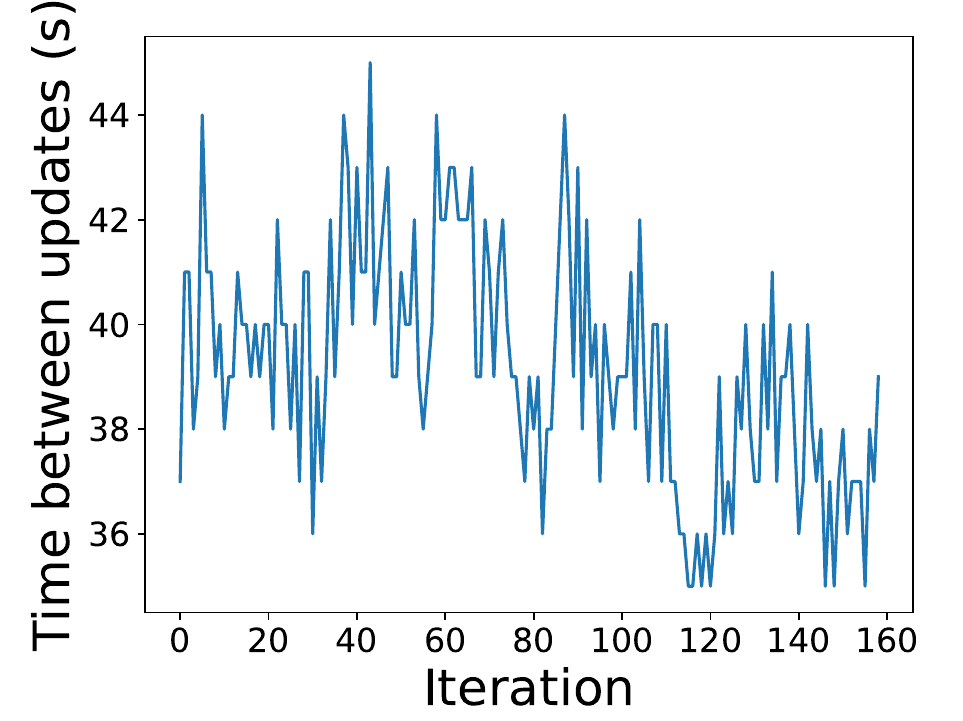}
        \caption{Time between updates for the entire training cycle.}
        \label{fig:global_times}
    \end{subfigure}
    \caption{Most nodes complete their training in under 2.5 seconds. The overall iteration time in BSP depends on the straggler nodes.}
    \label{fig:BSP_drawback11}
\end{figure}

\begin{figure}[t]
    \centering
    \begin{subfigure}{0.24\textwidth}
        \includegraphics[width=\textwidth]{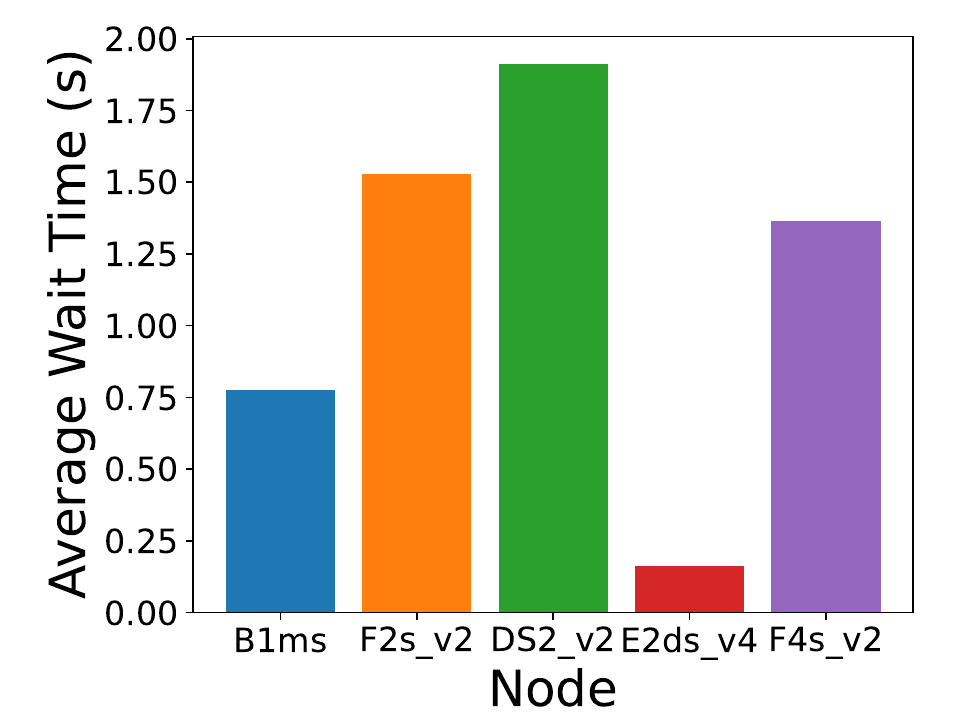}
        \caption{Wait times per node until gradients are pushed to PS.}
        \label{fig:node_waiting_times}
    \end{subfigure}
    \hfill
    \begin{subfigure}{0.24\textwidth}
        \includegraphics[width=\textwidth]{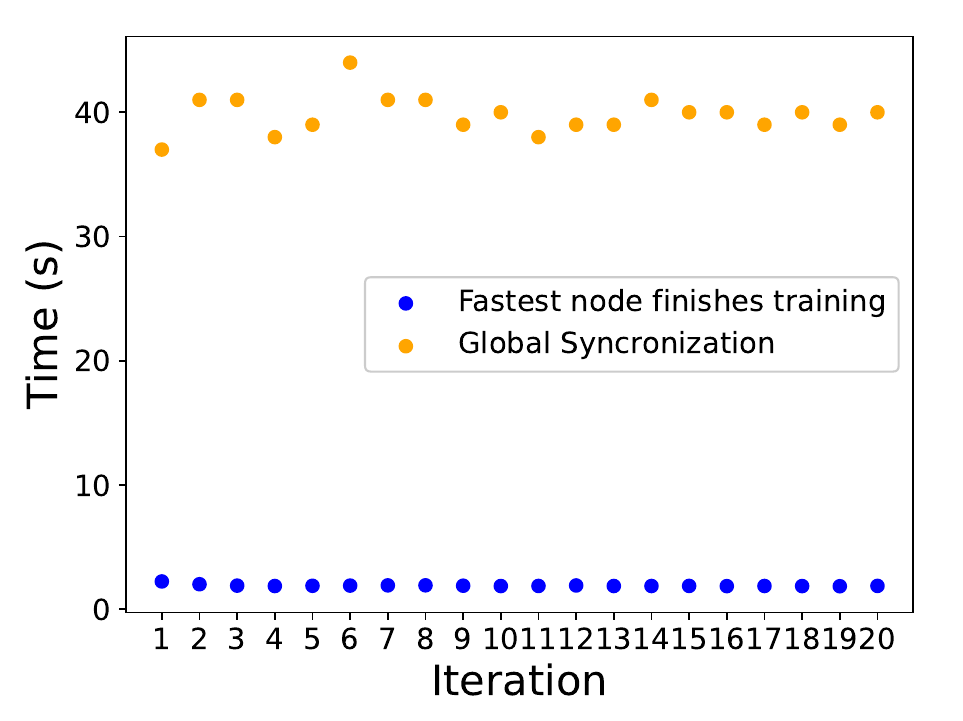}
        \caption{Waiting time for the fastest node in BSP (DS2\_v2 - as in Table \ref{table:4}) until gradients are pushed to PS.}
        \label{fig:global_times}
    \end{subfigure}
    \caption{Wastage of compute due to waiting for stragglers in BSP.}
    \label{fig:BSP}
\end{figure}

The major drawback of BSP is that it suffers from system heterogeneity, where workers have computational capabilities, and thus overall time is dependent and limited by the slowest worker or straggler. The synchronization time interval causes a slowdown in the overall training process and represents a wastage of computational resources. The BSP model's reliance on barriers for synchronization can lead to blocking. This is evident in Fig. \ref{fig:BSP_drawback11} and \ref{fig:BSP}, where all updates are being included indiscriminately even at the cost of wasting computational resources. Independence or asynchrony to workers is hence a desirable quality since it limits the excessive communication between the PS and the worker, however, there is a need to balance this independence with quality updates to the PS.

ASP and SSP have significant drawbacks despite their efficiency gains. As discussed in \ref{sec:asp}, ASP leads to more efficient use of compute but also introduces inconsistencies such as parameter staleness and non-deterministic convergence~\cite{dean2012large}. SSP imposes a controlled delay in synchronization, attempting to find a between parallelism and synchronization overhead. However, this leads to a lower convergence rate and the need to tune the staleness threshold $\delta$ for optimal performance and to manage the impact of straggler nodes. SelSync provides the concept of identifying changes in the gradients via relative gradient change, but as seen in \ref{sec:selsync}, this method is not useful. On the other hand, test loss, a metric we focus on, provides a more direct measure of model generalization performance on unseen data~\cite{ng1997preventing}. Unlike relative gradient change, it reflects the actual impact of training decisions on the final model's ability to handle real-world scenarios.

Nodes might generate updates of varying significance, based on their respective local data. If all updates are indiscriminately aggregated, it may lead to an increase in noise and redundancy in the model, implying decreased convergence rates. Therefore, identifying and selecting only major updates that contribute substantially to the global model optimizes the efficiency and effectiveness of DML.

\subsection{Need for a custom SGD algorithm}
SyncSGD ensures all workers operate on the most recent model parameters before each update, leading to theoretically better convergence compared to asynchronous methods. However, SyncSGD requires waiting for all workers to finish local training before proceeding to the next update -- which significantly impacts training speed, especially in heterogeneous clusters. AsyncSGD, on the other hand, allows workers to update the global model independently, eliminating straggler wait times and achieving high hardware efficiency by fully utilizing computational resources. Limitations of the same have already been discussed in \ref{sec:asp}. Therefore, an optimal SGD for edge-DML is needed to evaluate the effectiveness of the gradients in aiding the global model's convergence.

\subsection{Need for custom dataset sizes per worker}
Straggler nodes create a bottleneck in training with static dataset sizes under current synchronization paradigms. Stragglers themselves are often created due to hardware degradation or data accumulation (such as, activation weights) in ML~\cite{dai2018toward}. Using identical dataset sizes across nodes, regardless of their computational power, causes slower nodes to fall behind in training, consequently resulting in the issues detailed in \ref{sec:major}.

Custom dataset sizes for nodes in DML present a promising solution to these issues. By tailoring dataset sizes to the capabilities of individual nodes, more efficient use of available resources can be ensured, reducing idle times and improving throughput. This also enhances scalability by balancing the load across nodes. Additionally, careful adjustment of batch sizes can maintain model consistency and convergence rates, potentially achieving higher accuracy than traditional methods.

\textbf{Summary.} We need a mechanism to identify and transmit only the most significant updates, reducing communication overhead. We also need a new SGD algorithm that considers the effectiveness of gradients for convergence, potentially combining the benefits of both approaches. Furthermore, we need to try to avoid the problem of straggler nodes and allow them to actively participate in the training process. This thereby avoids the issue of stale gradients as well by dynamically setting dataset sizes for workers.

\section{System Design of \proj}

\proj revolves around four major key components:
\begin{itemize}
    \item Allocation of a dynamic batch size to each node via binary search, maximizing compute utilization while considering the issue of straggler nodes and stale gradients.
    \item Utilization of a probabilistic algorithm to identify significant improvements in model generalization capability.
    
\begin{figure*}[htb]
    \centering
    \begin{minipage}[b]{0.49\textwidth}
        \centering
        \includegraphics[width=\textwidth]{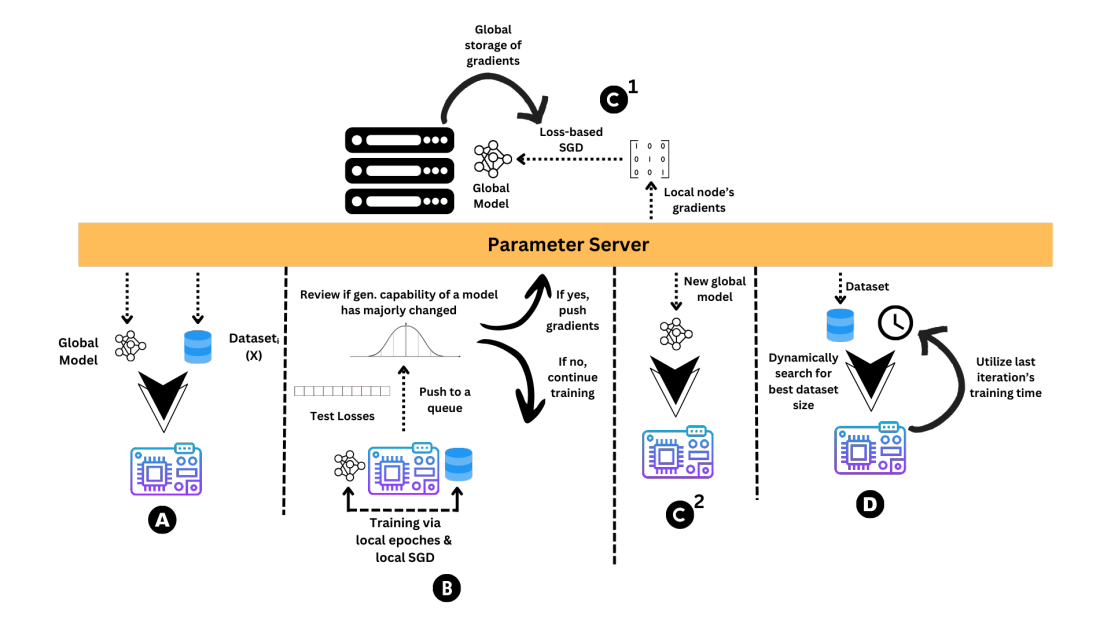}
        \caption{Workflow of the \proj framework; (a) Initialization, (b) Identification of a major change of the generalization capability for a local model, (c$^{1}$) Workflow for the loss-based SGD on the PS, (c$^{2}$) Refreshing a local worker's model with the global model when gradients were sent, (d) Using a dual-binary search algorithm to dynamically balance training data between stragglers and other nodes.}
        \label{fig:overall}
    \end{minipage}
    \hspace{0.00\textwidth} 
    \begin{minipage}[b]{0.49\textwidth}
        \centering
        \includegraphics[width=\textwidth]{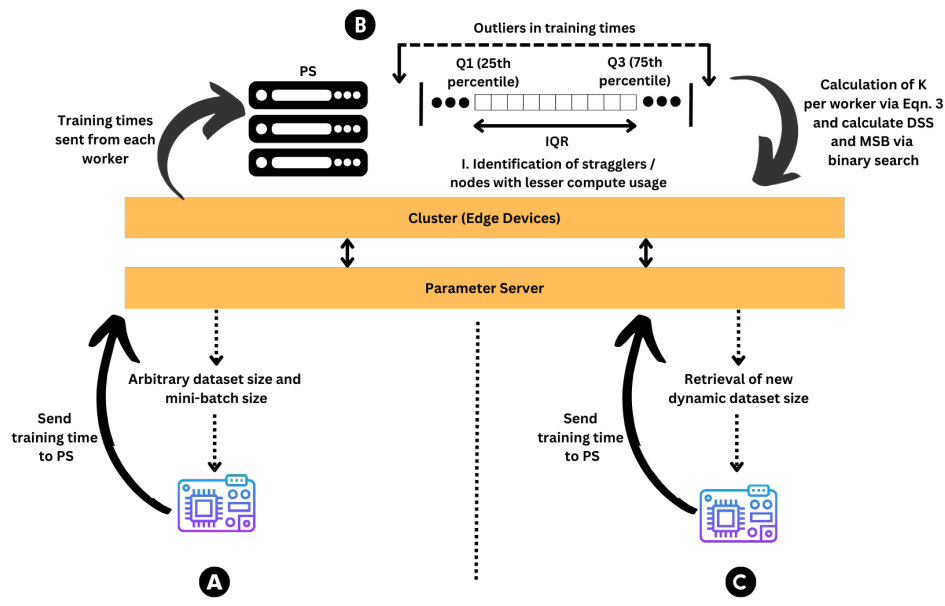}
        \caption{Workflow of calculating the dynamic dataset and mini-batch size via dual-binary search -- (a) Initialization, (b) Training time outlier identification \& dynamic batch size and mini-batch-size searching via binary search, (c) Prefetching; complexity = ~$O(lg N)$.}
        \label{fig:binary_search}
    \end{minipage}
\end{figure*}

    \item Creation of a custom loss-based SGD algorithm for the PS, to weigh a worker's gradient according to its generalization capability and effectiveness to contribute to a global model's convergence.
    \item Integration of optimization techniques asynchronously (prefetching and model compression) to further enhance communication efficiency and convergence speed.
\end{itemize}



As visualized in Fig. \ref{fig:overall}, these major components work seamlessly through a simple and effective workflow in four steps as defined below. We refer to the nodes where training takes place collectively as workers.
\begin{itemize}
    \item To begin with, the PS sets a static arbitrary dataset size and mini-batch size for all workers to train on, ensuring that the dataset size and fitted model do not exceed the memory limits of each worker (or in this case, the worker with least memory). To accommodate the memory constraints of all worker devices, the PS leverages mixed-precision training~\cite{micikevicius2017mixed}, enabling faster training and the deployment of smaller models.
    \item The PS distributes the allocated dataset to each worker. Workers perform local SGD on their assigned datasets. Using the algorithm \textit{HermesGUP}, we identify if the current training cycle has seen a major positive change in the generalization capability of the model.
    \item Upon retrieving the gradients from a worker, the PS utilizes a loss-based SGD algorithm to obtain a global model. This focuses the PS to update on the most informative gradients that contribute to better generalization. This global model is then sent back to the worker. Note that this process of aggregation is asynchronous.
    \item The PS also asynchronously monitors the performance of the workers and their training times. Due to certain nodes slowing down over time, we attempt to redefine the dataset size and mini-batch size for each worker, to avoid the worker being a straggler. As a result, the PS attempts to prefetch the necessary data to the worker thereby avoiding the creation of stragglers, while also allowing workers which have a large amount of compute power to contribute more towards training.
\end{itemize}
By following this efficient workflow, \proj dynamically adapts to the capabilities of individual worker nodes, focusing communication and computation on significant improvements in model generalization. This leads to faster convergence times and efficient resource utilization in DML training.

\subsection{Dynamic dataset and mini-batch sizes via Binary Search}
As stated in our motivation, several current frameworks often bottleneck training due to limitations imposed by straggler nodes~\cite{verbraeken2020survey}. A static dataset size forces straggler nodes to process the same amount of data as faster nodes, leading to significant slowdowns that impact the overall training time. To combat this, our approach proposes a dynamic dataset size allocation for each worker node using binary search. By analyzing the flow of training in ML, training time can be broadly modeled as a function of three factors: the number of local epochs $E$, the amount of training data sent to the worker $DSS$, and the mini-batch size used for training $MBS$. The training time also has a component $K$ that is uniquely defined for each worker node, which is technically defined as the computation time for the node to calculate the loss and gradients per mini-batch. This can be defined in terms of a $\theta$ time-complexity:
\begin{equation}
\scriptsize
    t_{train} = \Theta(E * \frac{DSS}{MBS}) = K * E * \frac{DSS}{MBS}
    \label{eq3}
\end{equation}

Based on an arbitrary $DSS$ and $MBS$ set during the initialization of the framework, provided the dataset size and model size fits in memory, based on the worker with the least memory, we store the training times of each worker for that $DSS$. As a result, we notice the outliers in these training times via quartiles~\cite{backstrom2019mindthestep}. By using the concept of box plots and Q1, Q3, and inter-quartile range (IQR), we can accurately identify outliers by checking if $t_{train} \notin [Q1-1.5*IQR, Q3+1.5*IQR]$. Once we notice outliers in these training times, which are either straggler nodes or nodes where the compute power has not been efficiently utilized, we begin calculating the value $K$, based on the initial run.

Stragglers tend to provide a significant issue of providing stale gradients, especially since \proj is based on asynchrony. To combat this, we base our framework on the fact that stragglers must actively participate in the training process, regardless of the dataset size and avoid sending stale gradients. As a result, we need these nodes to have their training times to be around the training times of majority of the cluster, which we define as the median of the training times $t_{median}$. Given our recently calculated $K$, we use a double (or dual) binary-search to estimate $DSS$ and $MBS$ for the training time $t_{median}$. This approach ensures efficient resource utilization across the heterogeneous edge cluster. In this context, $DSS$'s domain varies from $[0, len(dataset)]$, provided the size of the dataset and model fits in memory, while we restrict $MBS$'s domain to be commonly used mini-batch sizes, which are of the powers of 2~\cite{perrone2019optimal}, such as $[2, 4, 8, 16 ... 256].$ Thus, we effectively define an algorithm which is $O(lg N *lg K)$, where $N$ is the size of the dataset and $K$ is the number of mini-batch sizes we consider. Since K is a small number, we can approximate our running time to be $O(lg N)$, which is considerably faster than benchmarking workers, as seen in EBSP.
A detailed workflow can be viewed in Fig. \ref{fig:binary_search}. Additionally, \proj implements data prefetching by loading data into the cache or memory of a worker before the start of the next iteration to optimize data retrieval time based on the allocated dataset size and mini-batch size for each node, further streamlining the training process, reducing wait time and improving performance.

\subsection{Identification of major changes to the generalization capabilities of a model}
\proj explores the limitations seen in previous works in \ref{sec:moti} and we conclude that test loss, evaluated on a separate dataset not used for training (as shown in \ref{sec:exp}), offers a more robust measure of generalization capability. This metric provides a more reliable indicator of how well the model performs on unseen data, ultimately reflecting the true goal of achieving good generalization~\cite{recht2018cifar}. As a result, we begin by defining a hyper-parameter $w$ that explores the last $w$ test losses on the worker.

\subsubsection{Queue to explore past iterations}
\begin{figure}[h]
\centering
    \includegraphics[width=7cm]{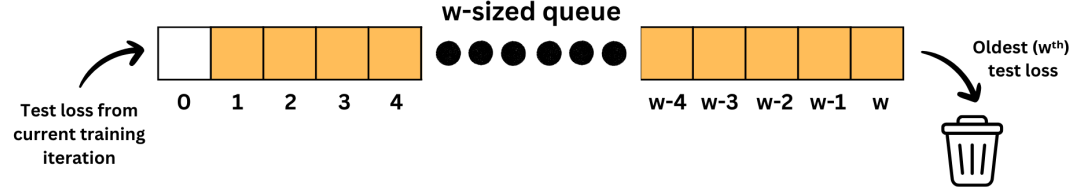}
    \caption{Proposed queue to store previous test losses.}
    \label{fig:queue}
\end{figure}

\proj involves leveraging a window-based approach to analyze recent test loss trends. We employ a queue data structure to store the past $w$ test losses on each worker, where $w$ is a hyper-parameter representing the queue size, as seen in Fig. \ref{fig:queue}. New test losses are appended to the queue, and the oldest value is discarded if the queue reaches its size $w$. We only consider the recent performance history of the model because we capture the recent magnitude of changes in generalization capability and hence, ignore the fluctuations and temporary increases in loss in older iterations (if we consider storing all the test losses). Furthermore, this approach reduces sensitivity to random fluctuations and noise in the loss function in past iterations. By focusing on recent trends, we are less likely to be misled by temporary spikes or dips in test loss. From this queue, we consider the current iteration of a worker and hence, utilize \textit{HermesGUP} to verify if there has been a major change in the generalization capability of the worker's model.

\subsubsection{HermesGUP - Statistically defining a major change in the generalization capability of a model}
HermesGUP leverages the concept of the $z$-score to evaluate how much a current test loss ($x$) deviates from the average performance within the recent window of size $w$. In the context of DML training, test loss can be viewed as a function of many factors, including model parameters, optimizer updates, and inherent randomness in the training process~\cite{zhang2021understanding}. While the individual factors might not be normally distributed, the Central Limit Theorem~\cite{kwak2017central} suggests that the sum of their effects (reflected in the test loss) can be approximated by a normal distribution, especially with a sufficiently large number of training iterations. Test loss, as a metric itself, can be viewed as an independent variable because dependence between successive losses is seen to be non-existent. Furthermore, the queue provides a more stable estimate of the underlying distribution of test losses. 

\begin{figure}[h]
\centering
    \includegraphics[width=6cm]{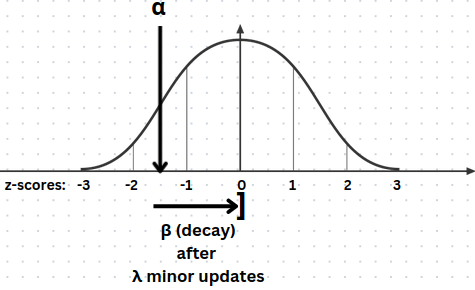}
    \caption{Description of proposed hyper-parameters $\alpha, \beta, \lambda$ for \proj.}
    \label{fig:hyper}
\end{figure}

The z-score essentially standardizes a data point by expressing its deviation from the mean in terms of standard deviations, which can be directly represented on a normal distribution, as seen in Fig. \ref{fig:hyper}. The $z$-score is defined below, where $\mu$ represents the expected mean of the losses in the queue and $\sigma$ represents the standard deviation of the losses in the queue.
\begin{equation}
\scriptsize
    z = (x - \mu) / \sigma
\end{equation} 
Since the normal distribution is bell-shaped and centered around the $\mu$, negative z-scores correspond to test loss values that are lower than the average recent performance captured by the window. As a result, a more negative z-score indicates a greater deviation from the mean and suggests a higher probability of this improvement being statistically significant. Thus, our algorithm considers only negative z-scores to be an improvement to the global model. However, not all negative z-scores are equally significant. To address this, we introduce a hyper-parameter called $\alpha$. $\alpha$ defines the minimum acceptable z-score for a test loss to be considered a statistically significant improvement (e.g. -1.3). This threshold ensures a high level of confidence that the current test loss represents a true improvement in generalization capability. Hence, $\alpha$ would ensure that the generalization capability of the model has statistically improved, making the worker push its gradients to the PS for aggregation.

\subsubsection{Dynamic nature of $\alpha$}
To avoid excessive communication overhead, \proj avoids major updates in the initial phases of training as the model learns basic patterns. However, to accelerate the global model reaching convergence, our framework must consider major updates as local worker models converge to a minimum test loss~\cite{li2020federated}. To address this, we introduce a dynamic $\alpha$ that adjusts based on the number of training iterations since the last major update, defined by the hyper-parameter $\lambda$. A highly negative value $\alpha$ indicates that the generalization capability has significantly changed from the last gradient push to the PS, which is suitable to avoid major updates in the initial phases of training. As the local model reaches convergence, our framework attempts to identify crucial updates as we reach a loss close to convergence. To define "closeness to convergence", we state that number of iterations ($N_{iter}$) since the last push to the PS. Thus, if $N_{iter} >= \lambda$, we decay $\alpha$ by another hyper-parameter $\beta$; else, we continue using the same alpha. The dynamic alpha mechanism balances the need to reduce communication overhead in earlier stages with the requirement to identify smaller but crucial improvements near convergence. This is shown in Algorithm~\ref{algo:gup}. 

\IncMargin{1em}
\begin{algorithm}[htbp]
\SetKwData{Left}{left}
\SetKwData{This}{this}
\SetKwData{Up}{up}
\SetKwFunction{Union}{Union}
\SetKwFunction{FindCompress}{FindCompress}
\SetKwComment{comment}{\#}{}
\BlankLine
\scriptsize
$Q \gets $ last $w$ test losses computed by a worker, calculate $\mu$ and $\sigma$ from this queue \\
$N_{iter} \gets 0$, this refers to number of iterations from last major update. \\
\While{worker is training}{
    \textit{complete one local training iteration; let $x$ be the test loss on current iteration} \\
    compute $z_{x} \gets x - \mu / \sigma$ \\
    \If{$z_{x} <= \alpha$} {
        push gradients to PS, wait for global model and dataset \\
        $N_{iter} \gets 0$
    }
    \Else {
        $N_{iter} \gets N_{iter} + 1$ \\
    }

    \If{$N_{iter} >= \lambda$} {
        $\alpha \gets \alpha * (1 - \beta)$ \\
    } 
}
\caption{\textit{HermesGUP}: Gradient Update Push.}
\label{algo:gup}
\end{algorithm}

\subsection{Loss-based SGD at the PS}
As major updates are pushed to the PS, we propose a novel approach called ``loss-weighted aggregation" for gradients. This method prioritizes updates that signify significant progress toward convergence, thereby reducing unnecessary communication and accelerating the training process. To achieve this, we introduce test loss evaluations. The PS computes a temporary local model using only the most recent worker's gradients. By comparing the test losses of the local model and the global model, the PS can determine whose gradients can contribute more to reducing the overall loss. This is illustrated in Algorithm~\ref{algo:sgd}. Mathematically, loss-based SGD is shown in Equations~\ref{eq5} and \ref{eq6}, where $w_{0}$ refers to the freshly initialized parameters of the model, $T_{w}$, $G$ represents the test loss and gradients sent by the worker, which is defined as sum of gradients (from all previous iterations of local-SGD on the worker -- which represents the worker's local model from the initial model with parameters $w_{0}$), $T_{g}$, $\varsigma$ represents the test loss and sum of gradients (from all previous iterations of gradient aggregation on the global model -- which represents the global model from the initial model with parameters $w_{0}$):
\begin{equation}
\scriptsize
    W_{1} = 1 / T_{w} ; W_{2} = 1 / T_{g}
    \label{eq5}
\end{equation}
\begin{equation}
\scriptsize
    w_{global} = w_{0} - \eta\frac{W_{1}G+W_{2}\varsigma}{W_{1} + W_{2}} 
    \label{eq6}
\end{equation}

As seen in Equations~\ref{eq5} and \ref{eq6}, the weights for aggregation are the reciprocal of the losses and hence, allow the global model to tend to gradients that contribute more to convergence. This process of computation of $G$ on the worker is not very intensive, since we deal with primitive addition and subtraction operators only. Therefore, this loss-weighted aggregation technique complements the \textit{HermesGUP} and allows \proj to reach convergence faster.

\IncMargin{1em}
\begin{algorithm}[htbp]
\SetKwData{Left}{left}
\SetKwData{This}{this}
\SetKwData{Up}{up}
\SetKwFunction{Union}{Union}
\SetKwFunction{FindCompress}{FindCompress}
\SetKwComment{comment}{\#}{}
\BlankLine
\scriptsize
\textbf{Input:} $\eta$ learning rate, $G$ gradients from a worker, $M$ baseline global model with freshly initialized weights $w_{0}$, $x_{test}$ test dataset \\

\textbf{Worker-SGD:} $G$ at the worker must be defined as the following: \\
    \Indp$G \gets G + \epsilon$
    \tcp*{$\epsilon$ represents the gradients from the current training step. Essentially we are calculating the gradients to the current local model directly from the baseline model.} 
    \Indm
\textbf{Initial step at PS:} This step refers to when the first worker pushes their gradients to the PS. \\
\Indp
    $\varsigma \gets G$
    \tcp*{$\varsigma$ refers to a storage of global gradients}
    $w_{1} = w_{0} - \eta * \varsigma$ \\
    Evaluate the model with $x_{test}$ and store the test loss in $L$. \\
\Indm
\textbf{PS-SGD (after initial):} This step refers to when any worker pushes their gradients to the PS. \\
\Indp
    $w_{temp} = w_{0} - \eta * G$ \\
    Evaluate the model of weights $w_{temp}$ with $x_{test}$ and store the loss as $L_{temp}$. \\
    $W_{1} \gets 1/L$, $W_{2} \gets 1/L_{temp}$ \\
    $w_{global} = w_{0} - \eta * \frac{W_{1}\varsigma + W_{2}G}{W_{1} + W_{2}}$ \\
    Evaluate the model with weights $w_{global}$ with $x_{test}$ and store the loss as $L$. \\
    $\varsigma \gets \frac{W_{1}\varsigma + W_{2}G}{W_{1} + W_{2}}$ \\
\Indm
\caption{Loss-based SGD.}
\label{algo:sgd}
\end{algorithm}

\subsection{Prefetching and model compression}
Model compression, by reducing the model size and number of parameters, inherently lowers memory requirements. This allows workers with limited memory to train on larger effective datasets allocated by the dual-binary search algorithm. One approach involves utilizing mixed precision training, particularly utilizing the fp16 precision (16-bit floating-point precision) instead of the standard fp32 precision (32-bit). This reduces memory footprint by half while maintaining reasonable accuracy for many DML tasks~\cite{micikevicius2017mixed}. Furthermore, prefetching complements the dual-binary search algorithm by proactively fetching new training data while the worker processes the current batch. By keeping all workers engaged in parallel and avoiding straggler effects, prefetching contributes to faster overall training times. 

\begin{figure}[h]
    \includegraphics[width=\columnwidth]{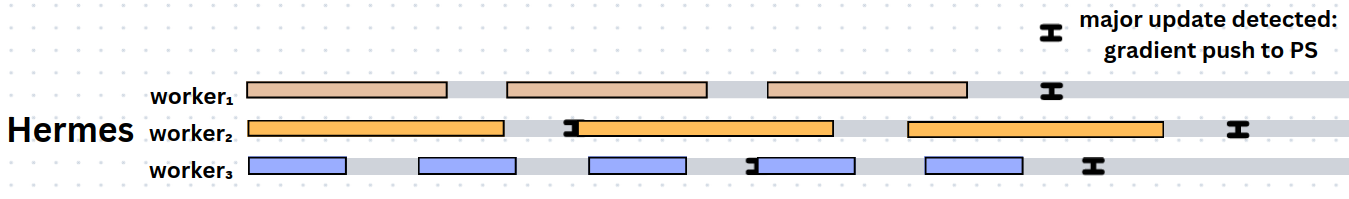}
    \caption{Training and communication time for \proj to highlight the difference between the setups in other state-of-the-art frameworks, similar to the worker configuration in Fig. \ref{fig:all_sota}.}
    \label{fig:setup}
\end{figure}

\textbf{Summary.} \proj hence tackles the challenges of communication overhead and slow convergence in DML training through a multi-pronged approach. Fig.~\ref{fig:setup} displays the barriers set up by \textit{HermesGUP} while presenting the training and communication times. Moreover, researchers and developers can actively train their custom models and datasets using this open-source framework.


\section{Evaluation of \proj}
The evaluation of \proj addresses the following questions:
\begin{itemize}
    \item Does \proj achieve faster convergence compared to state-of-the-art (SOTA) methods? Does it have a reduction in the training time?
    \item How does \proj's dual-binary search approach influence the issue of stragglers when compared to static allocation methods?
    \item How often do major updates occur in a particular node; does statistically defining an improvement in generalization capabilities work for a particular worker?
    \item Is there any significant difference when we attempt to change $\alpha$ and $\beta$ during training?
\end{itemize}

\subsection{Experimental testbed} \label{sec:exp}
\textit{Data.} We utilize the MNIST~\cite{lecun1998gradient} and CIFAR-10~\cite{krizhevsky2009learning} datasets, which each encompasses 60K images with 10 different labels. MNIST acts as an IID dataset, while CIFAR-10 acts as a non-IID dataset. The data is divided into fixed train (85\%), and test (15\%) sets, and all results are reported based on this split. While MNIST and CIFAR-10 are simple benchmarks, they are effective for measuring the impact of major changes, such as batch size variations, and are widely used in DML ~\cite{horvath2022naturalcompressiondistributeddeep}~\cite{George_Gurram_2020}. 

\textit{Models.} We employ a CNN for the MNIST dataset and downsize AlexNet~\cite{krizhevsky2012imagenet} for the CIFAR-10 dataset. Our CNN contains approximately 110K parameters and downsized AlexNet contains approximately 990K parameters. The diverse models and datasets used in our evaluation would prove to define the validity of \proj. 

\textit{Hyper-parameters.} For \proj, we define these universal hyper-parameters for these datasets. Unless otherwise specified, we set the following parameters as specified in Table~\ref{table:2}.
\begin{table}[h!]
\centering
\begin{tabular}{|l|l|}
\hline
\textit{Model} & \textit{Hyper-parameters} \\ \hline
\begin{tabular}[c]{@{}l@{}}CNN\\ (parameters: $110$K,\\ optimizer: SGD)\end{tabular} & \begin{tabular}[c]{@{}l@{}}$\eta = 0.1$,\\ patience = $25$, \\ $\lambda = 5$\end{tabular} \\ \hline
\begin{tabular}[c]{@{}l@{}}AlexNet\\ (parameters: $990$K,\\ optimizer: SGDM)\end{tabular} & \begin{tabular}[c]{@{}l@{}}$\eta = 0.001$,\\ patience = $10$, \\ momentum = $0.9$, \\ $\lambda = 15$\end{tabular} \\ \hline
\end{tabular}
\caption{Hyper-parameters utilized for training on \proj; $w$ is set to 10 in both models.}
\label{table:2}
\end{table}
\raggedbottom

\textit{Baselines.} We evaluate our framework against SOTA models, including BSP, ASP, SSP, and Elastic BSP with two major models using the same hyper-parameters (Table~\ref{table:2}).

\textit{Workers and PS.} We test \proj on a system of 12 diverse workers, specifications detailed in Table \ref{table:4}, and one PS, which holds a Intel i7-12700H with 64GB RAM. Communication is conducted by a custom-made ZeroMQ (ZMQ) API that sends and receives major variables between the workers and PS. ZMQ offers a high-performance asynchronous messaging framework, enabling efficient exchange of gradients, model updates, and control messages between workers and PS. Furthermore, we utilize Kafka and SFTP for dataset and model transfer respectively. Communication setups are consistent across all frameworks. 

\begin{table}[h!]
\centering
\begin{tabular}{||c c c||} 
 \hline
 \textit{Workers x Quantity} & \textit{CPU Cores (vCPU)} & \textit{RAM (GB)}\\ [0.5ex] 
 \hline\hline
 B1ms x 2 & 1 & 2 \\ 
 F2s\_v2 x 3 & 2 & 4 \\
 DS2\_v2 x 3 & 2 & 7 \\
 E2ds\_v4 x 2 & 2 & 16 \\
 F4s\_v2 x 2 & 4 & 8 \\ [1ex] 
 \hline
\end{tabular}
\caption{Experimental setup consisting of diverse memory-optimized and compute-optimized workers.}
\label{table:4}
\end{table}
\raggedbottom

\begin{table*}[]
\centering
\begin{tabular}{|c|c|c|c|c|c|c|c|c|}
\hline
\textit{Dataset} & \textit{Model} & \textit{Framework} & \textit{Iterations} & \textit{Time taken (min)} & $WI_{avg}$ & \textit{Conv. Acc.} & \textit{Avg. API Calls} & Speedup \\ \hline
\multirow{7}{*}{\textit{MNIST}} & \multirow{7}{*}{\textit{CNN}} & BSP & 9600 & 105.38 & 1.00 & 98.07\% & 18.9M & 1.00x \\ \cline{3-9} 
 &  & ASP & 15325 & 55.72 & 1.00 & 90.625\% & 24.6M & 1.89x \\ \cline{3-9} 
 &  & SSP ($s = 125$) & 20955 & 221.63 & 1.00 & 90.79\% & 29.6M & 0.47x \\ \cline{3-9} 
 &  & E-BSP ($R = 150$) & 3840 & 124.28 & 5.09 & 85.34\% & 15.5M & 0.84x \\ \cline{3-9} 
 &  & \textbf{Hermes ($\alpha = -0.9, \beta = 0.1$)} & \textbf{2200} & \textbf{11.2} & \textbf{7.41} & \textbf{97.71\%} & \textbf{11.2M} & \textbf{9.4x} \\ \cline{3-9} 
 &  & \textbf{Hermes ($\alpha = -1.3, \beta = 0.1$)} & \textbf{2830} & \textbf{10.55} & \textbf{7.90} & \textbf{98.02\%} & \textbf{13.5M} & \textbf{9.9x} \\ \cline{3-9} 
 &  & \textbf{Hermes ($\alpha = -1.6, \beta = 0.15$)} & \textbf{2250} & \textbf{7.97} & \textbf{8.70} & \textbf{97.82\%} & \textbf{11.4M} & \textbf{13.22x} \\ \hline
\multirow{5}{*}{\textit{CIFAR-10}} & \multirow{5}{*}{\textit{\begin{tabular}[c]{@{}c@{}}AlexNet\\ (downsized)\end{tabular}}} & BSP & 3360 & 292.58 & 1.00 & 51.73\% & 8.8M & 1.00x \\ \cline{3-9} 
 &  & ASP & 5320 & 137.49 & 1.00 & 40.56\% & 9.8M & 2.12x \\ \cline{3-9} 
 &  & SSP ($s = 125$) & 4780 & 183.82 & 1.00 & 44.27\% & 9.7M & 1.59x \\ \cline{3-9} 
 &  & E-BSP ($R = 150$)* & - & - & - & - & - & - \\ \cline{3-9} 
 &  & \textbf{Hermes ($\alpha = -1.6, \beta = 0.15$)} & \textbf{1470} & \textbf{43.39} & \textbf{9.12} & \textbf{53.27\%} & \textbf{8.2M} & \textbf{6.74x} \\ \hline
\end{tabular}
\caption{Performance across \proj, BSP, ASP, SSP, EBSP; Convergence Accuracy (Conv. Acc.), Avg. API Calls (recorded per million) and Speedup (for Time) compared to BSP. EBSP fails to converge very well in both models, due to the high compute power required for benchmarking, leading to several workers crashing in multiple simulations.}
\label{table:1}
\end{table*}
\raggedbottom

\begin{figure}[h]
    \centering
    \begin{subfigure}{0.24\textwidth}
        \includegraphics[width=\textwidth]{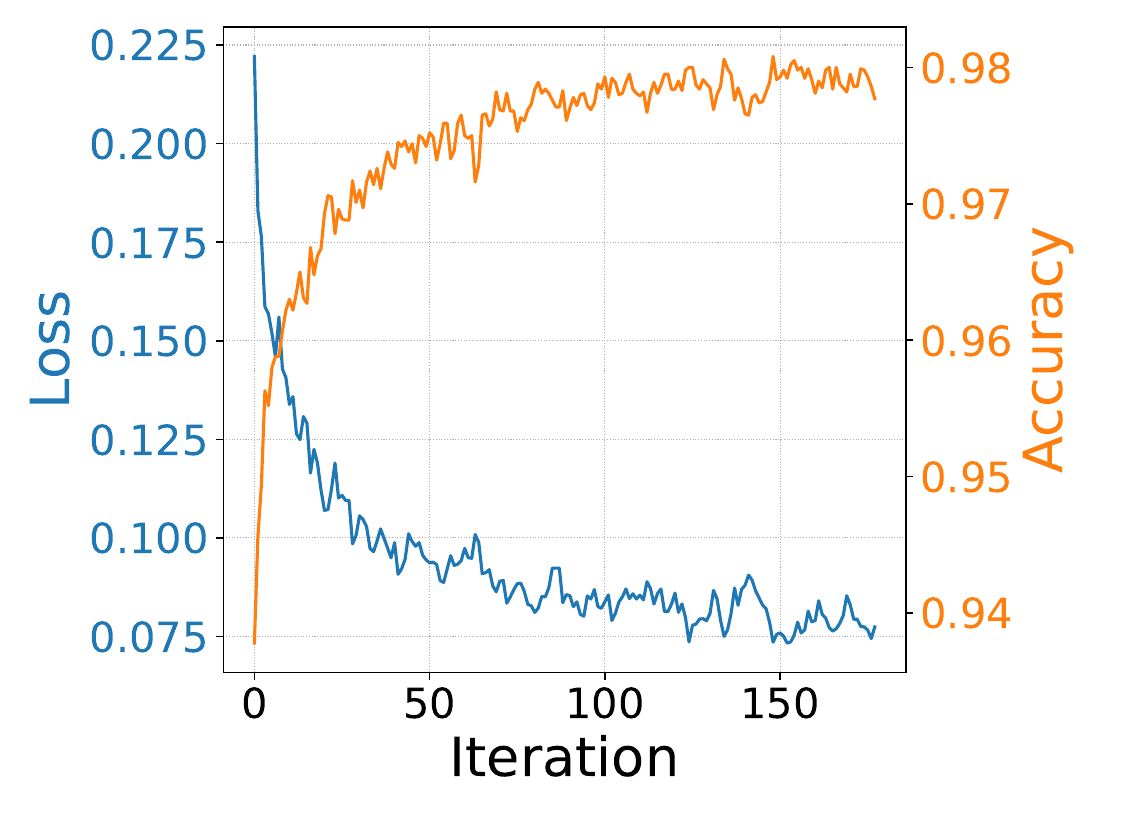}
        \caption{Test accuracy and test losses (MNIST, $\alpha = -1.3, \beta = 0.1$) for global model on PS.}
        \label{fig:test_acc}
    \end{subfigure}
    \hfill
    \begin{subfigure}{0.24\textwidth}
        \includegraphics[width=\textwidth]{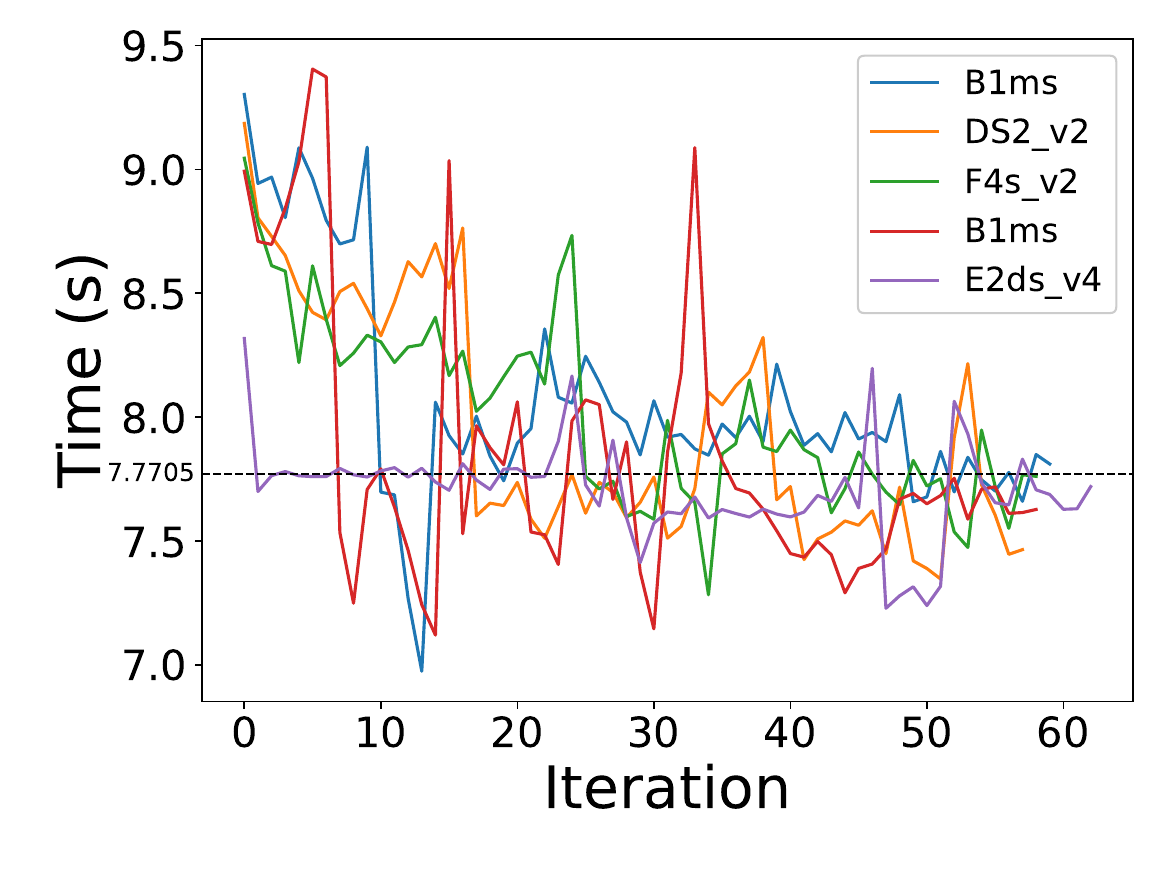}
        \caption{Stabilization of training times for all workers throughout the training process (MNIST).}
        \label{fig:training times}
    \end{subfigure}
    \caption{Performance of \proj by plotting test loss and accuracy of the global PS model, along with the training time of the workers.}
    \label{fig:BSP_drawback}
\end{figure}

\subsection{Performance of \proj against SOTA frameworks}
\label{performance}
The same cluster of 12 workers and 1 PS is also used to evaluate BSP, ASP, SSP, and EBSP. SSP is deployed with $s = 125$, allowing $s$ to be large enough to not be a frequent synchronization barrier, while still small enough to not let worker models diverge too far from the global model. \proj, on the other hand, is deployed with three different configurations of $(\alpha, \beta) \gets$ (-1.3, 0.1), (-1.6, 0.15), (-0.9, 0.1).

Fig. \ref{fig:test_acc} shows a plot of the global accuracy and overall training time using the above experimental setup and parameters. We observe that we obtain comparable accuracy (to BSP, refer Table~\ref{table:1}) in the context of MNIST (-0.05\%) and a positive accuracy change (+1.54\%) in the context of CIFAR-10. Regardless, we observe a major change in training time in comparison to \proj and BSP (7.97m, 1h45m). Furthermore, we notice that \proj has 62.1\% lesser communication activity than SSP, which is recorded via API calls in our experiments. This includes contacting the PS for the dataset, the model, global gradients and any other relevant information about other nodes. 

Fig. \ref{fig:training times} shows a plot of the training times of one worker from each of the five families listed in Table~\ref{table:4}. This also indicates the stabilization of the training times, allowing high-compute workers and stragglers to be around the training time of the majority of the cluster, which is indicated by the median of the training times, which is 7.705 seconds in this plot. 

We also introduce a novel metric to represent the level of independence per worker family, denoted by WI (Worker Independence - Equation~\ref{eq7}), which compares the number of local-SGD iterations a worker has done and the number of times the model has requested the PS for the global model.

\begin{equation}
\scriptsize
    WI_{worker} = \frac{iterations_{worker}}{req\_model_{global}}
    \label{eq7}
\end{equation}

This metric indicates the importance of local models achieving good progress via local SGD, even with limited communication and synchronization with the global model. From our evaluation, we find that \proj provides workers the most independence, decreasing excessive communication between the PS and the workers, hence, allowing the worker to concentrate on the local model's convergence and hence, contribute more significantly to the global model. $WI_{avg}$ for \proj (8.70) seems to be much higher than other SOTA frameworks (5.09, EBSP - highest among SOTA). 

All frameworks have been extensively tested and the mean results of three runs have been reported in Table \ref{table:1}. The total iterations that the framework has undergone, the total time to achieve a convergence accuracy, the average value of $WI$, convergence accuracy, and the total speedup in comparison to BSP have been reported. \proj achieved an impressive $13.22$x faster training time while maintaining accuracy comparable to BSP. This demonstrates \proj's effectiveness as a superior solution for DML training in resource-constrained edge environment. 

\subsection{Performance of nodes during dynamic dataset sizing}
\proj relies on combating the issue of straggler nodes by providing dynamic dataset sizes and mini-batch sizes for training. Furthermore, we experimentally prove that not only straggler nodes but also high-compute nodes are more participative in the process of training.

\begin{figure}[h]
    \centering
    \begin{subfigure}{0.24\textwidth}
        \includegraphics[width=\textwidth]{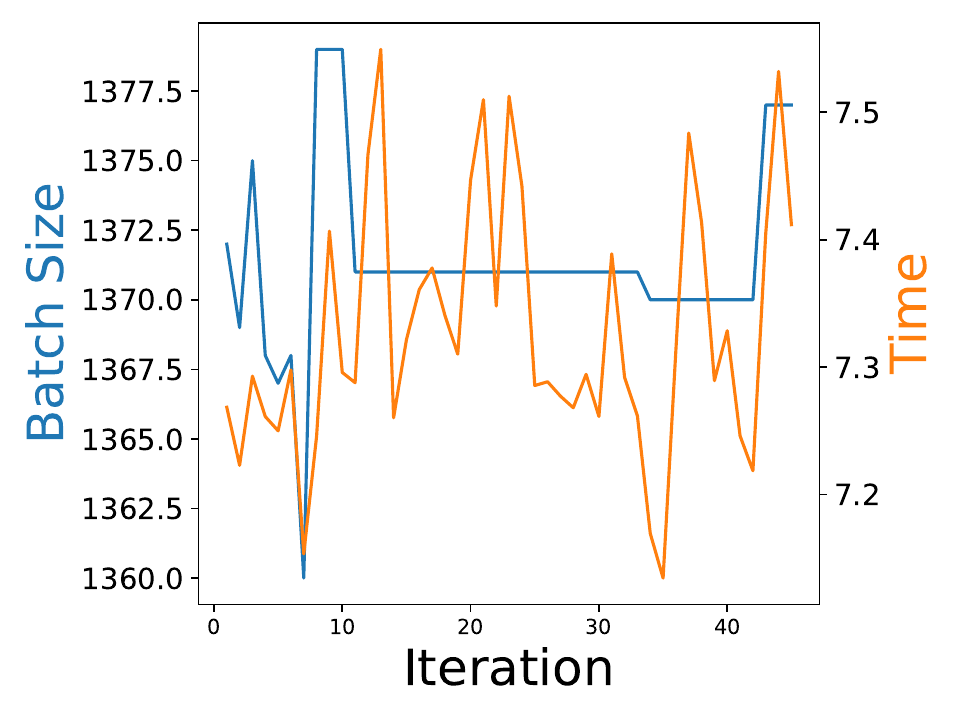}
        \caption{Dataset (batch) size sent to a worker and training times over the entire length of training; mini-batch size is seen to be consistently 16.}
        \label{fig:batch_time_all}
    \end{subfigure}
    \hfill
    \begin{subfigure}{0.24\textwidth}
        \includegraphics[width=\textwidth]{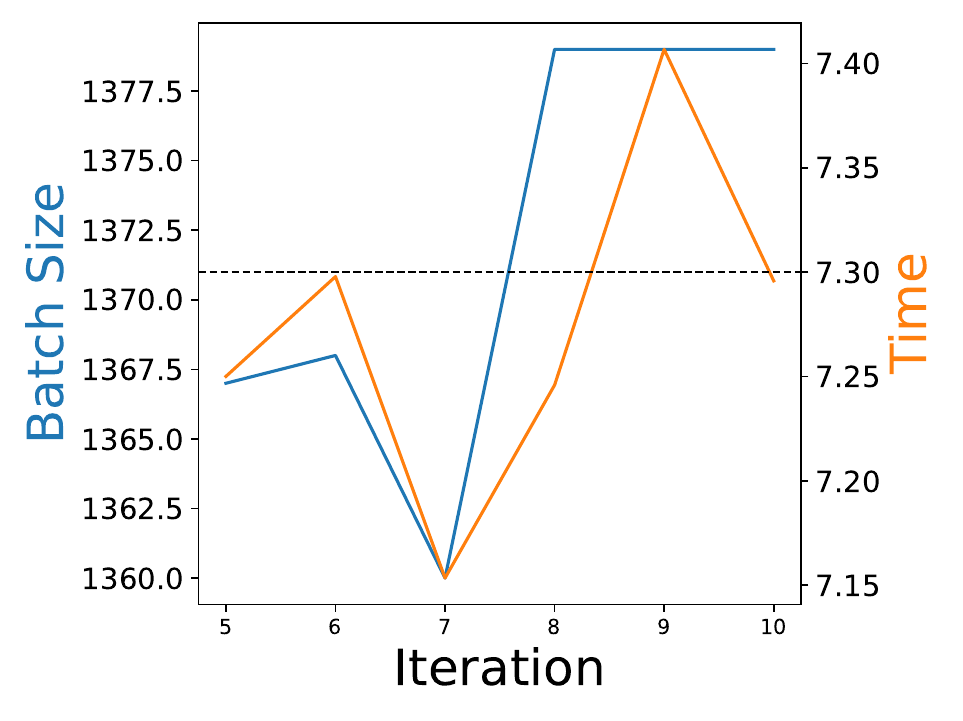}
        \caption{Dataset (batch) size sent to a worker and training times only from the 5th to 10th iteration; mini-batch size is seen to be consistently 16.}
        \label{fig:batch_time_5_10}
    \end{subfigure}
    \caption{Sensitivity of dataset size sent to the weakest worker versus the training time, initializing at 2500 images as the first dataset size and 16 as the mini-batch size.}
    \label{fig:dataset_sens}
\end{figure}

Fig. \ref{fig:batch_time_all} provides a comprehensive view of a node's changing dataset size sent by the PS during training, along with the training time per iteration, but we cannot visibly view the variations of training time appropriately. However, from Fig. \ref{fig:batch_time_5_10}, which looks at iterations 5 to 10, we can view how the variations in training time occur due to the changes in the dataset size, and how the worker's training time revolves around $7.3$ seconds, which is the median of the training times of nodes. The fluctuation around the median indicates the validity of Equation~\ref{eq3}. We verify that the PS adjusts the slowest node's training time to avoid the situation of being a straggler.
Furthermore, nodes with higher compute power tend to get more data from the PS as well. As a result, \proj demonstrates how this framework not only mitigates stragglers but also leverages high-compute nodes for faster training convergence in heterogeneous edge environments.

\subsection{Performance of HermesGUP and loss-based SGD during training}
\begin{figure}[h]
    \centering
    \includegraphics[width=5cm]{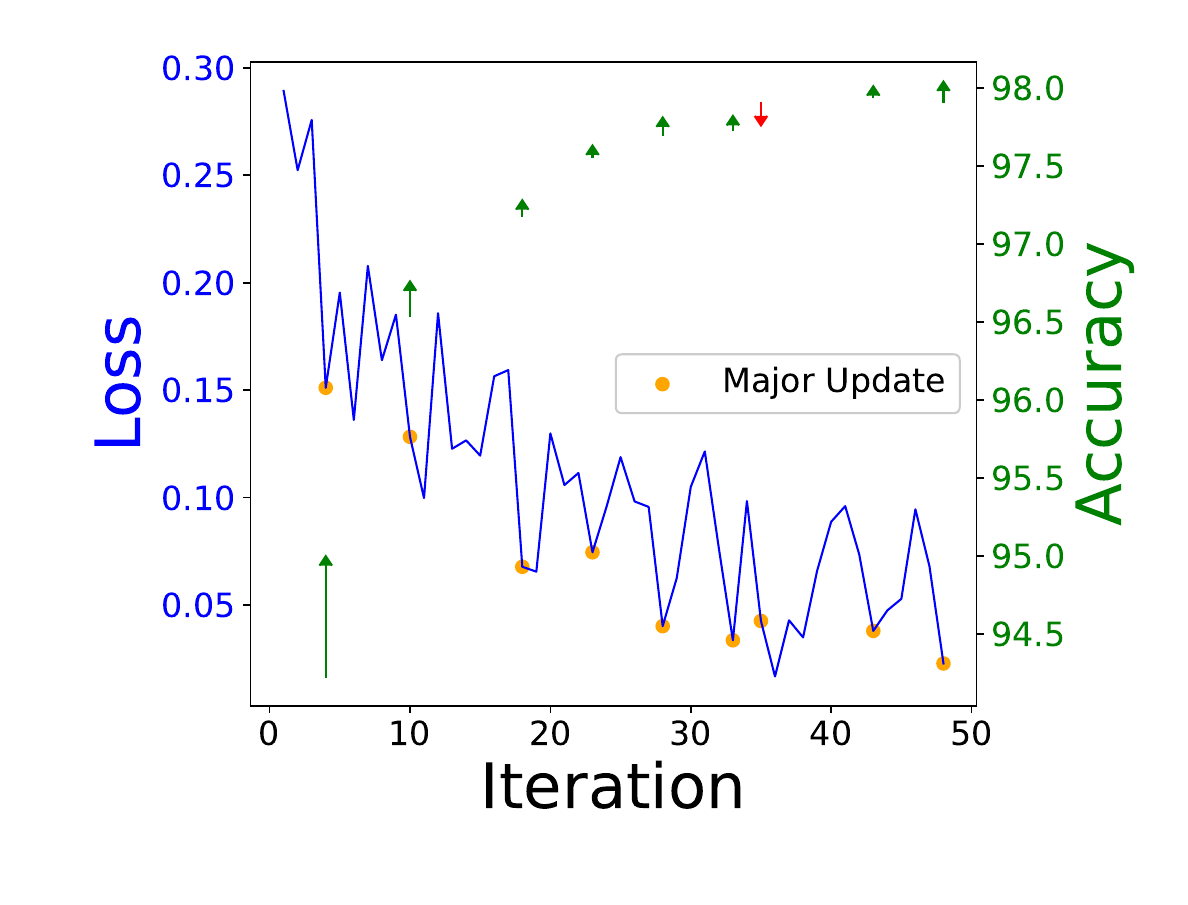}
    \caption{Test accuracy changes for the global model for each major update (highlighted with an orange dot) on a E2ds\_v2 running \proj on MNIST \& CNN.}
    \label{fig:accuracy_global}
\end{figure}

Fig. \ref{fig:accuracy_global} indicates the loss plots of a worker, marking the major updates by an orange dot. This plot proves the effectiveness of \textit{HermesGUP}, as these major updates occur only when a major change in test accuracy occurs, indicating a major change in the generalization capability of the model. This plot represents whenever the worker recognizes a major change in the generalization capability of its local model, along with loss-based SGD, the global model tends to converge faster and is guided in the right direction to the global minima.

Loss-based SGD, in this context, can be seen to be especially effective. As the local model reaches convergence and pushes major updates to the PS, the arrows represent the change of test accuracy on the global model after the aggregation of the pushed gradients. This SGD algorithm shows a steady increment (to convergence) as major updates are pushed to the global model. Therefore, \proj performs optimally and accounts for stale gradients (with their sub-optimal loss) to lead the global model to convergence.

\subsection{Significance of $\alpha$ and $\beta$ for training}
\begin{figure}[h]
    \centering
    \begin{subfigure}{0.24\textwidth}
        \includegraphics[width=\textwidth]{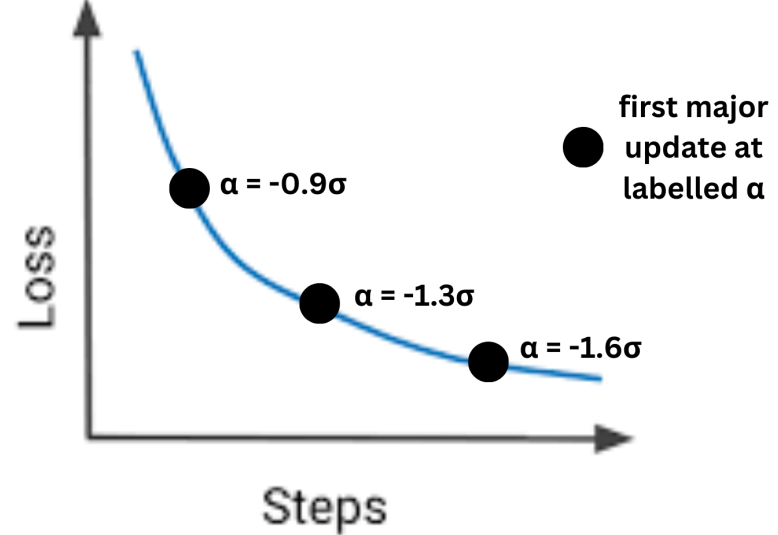}
        \caption{Loss vs Iteration; labelling points of major change depending on $\alpha$.}
        \label{fig:loss_iter_alpha}
    \end{subfigure}
    \hfill
    \begin{subfigure}{0.24\textwidth}
        \includegraphics[width=\textwidth]{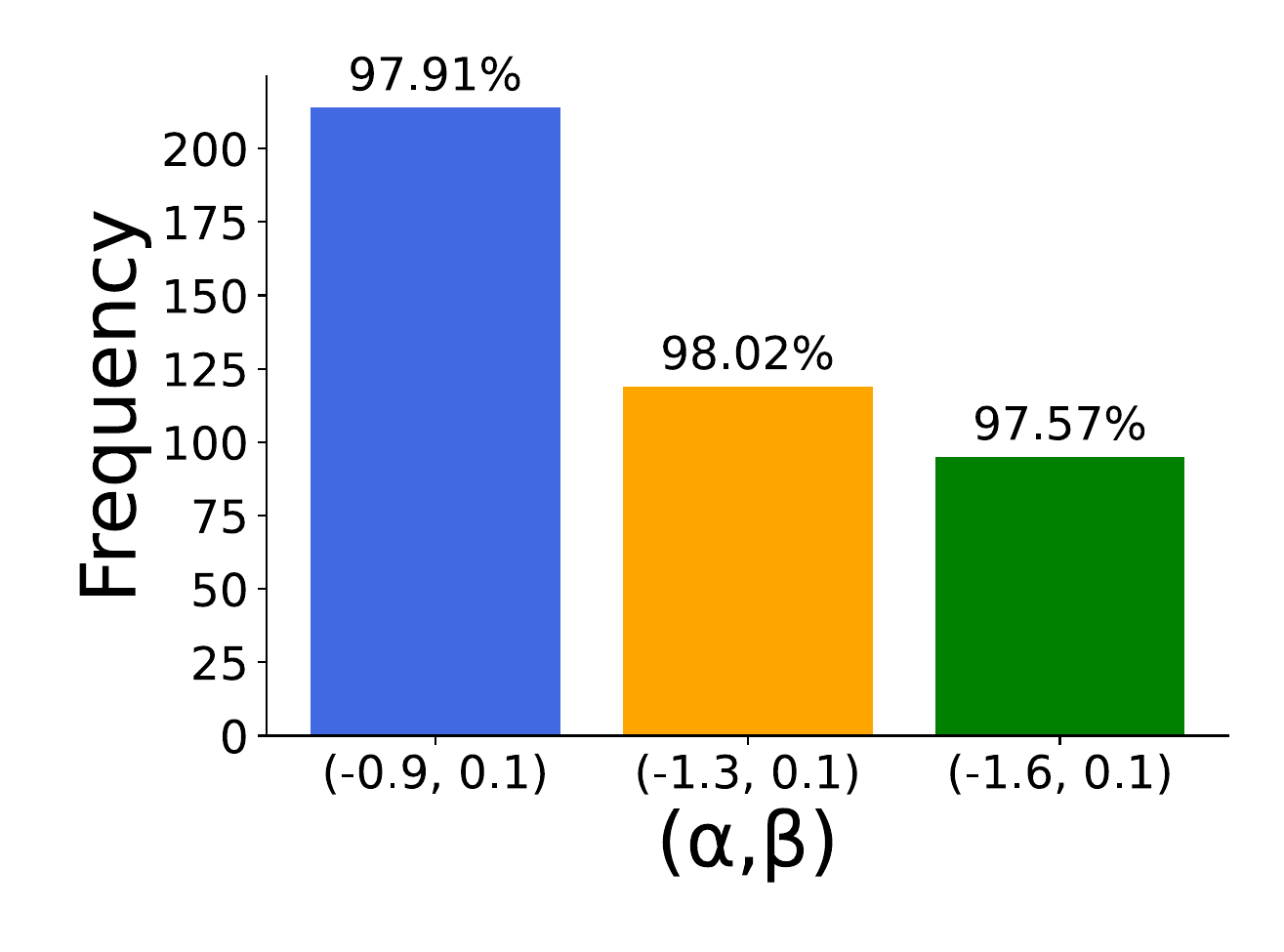}
        \caption{Frequency of major updates sent to PS with labelled conv. acc. for various values of $\alpha, \beta$.}
        \label{fig:update_freq}
    \end{subfigure}
    \caption{The importance and sensitivity of $\alpha$ and $\beta$ during training.}
    \label{fig:BSP_drawback}
\end{figure}

Hyper-parameters are often a crucial yet challenging aspect of ML algorithms. As a result, we evaluate various sets of hyper-parameters to review the importance of $\alpha$ and $\beta$ during training.
Fig. \ref{fig:update_freq} represents the frequency of pushing major updates to the PS with various $\alpha$, keeping $\beta$ constant. The final convergence accuracy is approximately the same with the maximum change in accuracy to be -0.45\%.  With a more negative $\alpha$, one can conclude that a major update is recognized when a model is statistically closer to the minima in the loss landscape, and hence, a lesser number of updates are required to achieve convergence. Statistically, utilization of $-1.3$ indicates that \textit{HermesGUP} classifies a test loss to be significant only if the probability of that test loss existing in the given queue (or distribution) is 9.68\%, whereas for $-1.6$, it is 5.48\% and for $-0.9$, it is 18.406\%. As a result, $\alpha$ is a sensitive hyper-parameter and the optimal $\alpha$ stands at how far away from the distribution of losses one wants to consider as a major change in the generalization capability. Fig.~\ref{fig:loss_iter_alpha} shows that $\alpha$ by itself is defined by how deep into the loss landscape \textit{HermesGUP} can recognize a major change. Picking a suitable $\alpha$ requires an analysis of the loss landscape. For example, more complex and dense datasets could utilize a more negative $\alpha$ for the local model to sufficiently explore the loss landscape.

The value of $\beta$ also holds significance as it tends to decay $\alpha$. This value should be dependent on how often one would like to have fine-tuning as the local model attempts to reach convergence. Experimentally, an optimal value for $\beta$ is seen to be between the range $[0.01, 0.15]$ for good performance.

\section{Discussions}
\subsection{Performance consistency of \proj in various clusters}
\proj is undergoing further independent testing across a wide range of nodes, including a custom cluster featuring multiple GPU-based nodes alongside fewer CPU-based devices, to assess its performance with large-scale models while maintaining node diversity for DML. These experiments show consistency with our results in Section \ref{performance}, confirming that \proj maintains stable performance without degradation.

As a result, \proj demonstrates strong adaptability when scaling up to diverse clusters, including GPU environments and larger models, highlighting its robustness in managing heterogeneous workloads. Dynamic batch sizes, paired with the efficient communication strategies, enables \proj to scale across clusters with varying bandwidth and computational capacities. In our original experiments and testing, we observe that \proj efficiently handles the inherent variability of nodes in real-world scenarios. Whether dealing with GPU-dense clusters or a mix of weaker CPU nodes, \proj successfully mitigates overhead by reducing insignificant updates and leveraging asynchronous communication. This ensures that both high-performance and lower-capacity workers are utilized optimally, stabilizing training times across diverse network environments.

\subsection{Hyperparameters \& batch sizes versus performance}
In \proj, hyperparameter tuning plays a crucial role in optimizing performance, particularly during distributed training. One key parameter, \(\alpha\), dynamically adjusts to control the communication of updates between worker nodes and the parameter server (PS). Early in the training phase, \(\alpha\) is set conservatively to reduce unnecessary communication overhead by only transmitting updates when they show statistically significant improvements. This prevents the system from sending updates too frequently, especially in the early stages when model parameters are still far from convergence. As training progresses and test loss stabilizes, \(\alpha\) is gradually relaxed via the decay factor \(\beta\), allowing \proj to capture more nuanced changes in the model that may still contribute to improved performance. Typically, \(\alpha\) ranges from \([-3, 0]\), with more negative values used to tighten the threshold for update transmission, ensuring updates occur only when the local model is closer to the loss minimum.

On the other hand, batch sizes, which are tailored to each worker's computational capacity, also have a significant impact on training performance. Weaker nodes are assigned smaller batches to avoid bottlenecks, while stronger nodes handle larger batches. This not only ensures that all workers contribute effectively but also helps balance the load across heterogeneous clusters, enhancing training efficiency and stability.

By dynamically adjusting these hyperparameters and batch sizes, \proj effectively manages the trade-offs between communication overhead and training performance, ensuring scalability across a variety of cluster environments.

\subsection{Future Work}
Our future work will involve conducting ablation studies to analyze which components of \proj contribute most significantly to its overall efficiency. These studies will help isolate the impact of key elements such as dynamic batching, asynchronous communication, and selective update transmission. By identifying the most critical factors, we can further optimize \proj and improve its performance.

Another promising area for future research is the potential integration of distributed schedulers with \proj. In this context, distributed schedulers could play a supportive role in coordinating across clusters by dynamically managing heterogeneous workloads and allocating resources based on demand. This would ensure efficient infrastructure management, allowing DML tasks to be distributed across multiple environments seamlessly. In such an arrangement, distributed schedulers would function above \proj, overseeing infrastructure reliability and resource allocation, while \proj remains focused on optimizing the DML training process at the framework level.

\section{Conclusion}
This paper introduces \proj, a novel framework designed to address the challenges of training DML models in resource-constrained edge environments. \proj tackles heterogeneous worker nodes and stragglers by employing a dynamic workload allocation strategy with dynamic dataset and mini-batch size allocation. This allows \proj to leverage the varying capabilities of individual worker nodes while mitigating the negative impact of stragglers on overall training time. The evaluation of \proj on a heterogeneous cluster with MNIST and CIFAR-10 datasets demonstrates that \proj achieves significant improvements in training speed ($13.22$x) and communication activity compared to SOTA frameworks while maintaining comparable or slightly better convergence accuracy. In conclusion, with the least number of iterations, \proj has proven to achieve similar or higher convergence accuracy to BSP, while significantly accelerating DML at the edge.

\section{Acknowledgment}
This work uses resources in the Data, Systems and HPC (DaSH) Lab at BITS Pilani, KK Birla Goa Campus, India. The lab resources are sponsored by SERB, Govt. of India Start-up Research Grant - SRG/2023/002445 and additional grants in BITS Pilani - BBF/BITS(G)/FY2022-23/BCPS-123/24-25/R1, GOA/ACG/2022-2023/Oct/11, and BITS CDRF - C1/23/173. The authors would also like to thank Imaad Momin, Palash Gupta and Kunal Mishra for their contribution towards this work.

\bibliographystyle{IEEEtran}
{\footnotesize
\bibliography{ refs.bib}}

\end{document}